\documentclass[conference]{IEEEtran}
\IEEEoverridecommandlockouts

\usepackage{adjustbox}
\usepackage{amsmath,amssymb,amsfonts}
\usepackage{algorithmic}
\usepackage{array}
\usepackage{booktabs}
\usepackage{cite}
\usepackage{enumitem}
\usepackage{graphicx}
\usepackage{multirow}
\usepackage{subcaption}
\usepackage{svg}
\usepackage{textcomp}
\usepackage{xcolor}

\DeclareUnicodeCharacter{0307}{\textasciitilde}

\usepackage{xspace}
\usepackage{graphicx}
\usepackage{ifthen}
\usepackage[normalem]{ulem} 
\usepackage{xcolor,colortbl}
\usepackage{hyperref}

\newboolean{showedits}
\setboolean{showedits}{true} 
\ifthenelse{\boolean{showedits}}
{
	\newcommand{\del}[1]{\textcolor{red}{\sout{#1}}} 
	\newcommand{\nbe}[3]{
		{\colorbox{#3}{\bfseries\sffamily\scriptsize\textcolor{white}{#1}}}
		{\textcolor{#3}{\sf\small$\blacktriangleright$\textit{#2}$\blacktriangleleft$}}}
}{
	\newcommand{\del}[1]{} 
	
	\newcommand{\nbe}[3]{}
}

\newboolean{showcomments}
\setboolean{showcomments}{true} 
\newcommand{\id}[1]{$-$Id: scgPaper.tex 32478 2010-04-29 09:11:32Z oscar $-$}

\ifthenelse{\boolean{showcomments}}
 {
 	\newcommand{\nbc}[3]{
 		{\colorbox{#3}{\bfseries\sffamily\scriptsize\textcolor{white}{#1}}}
		{\textcolor{#3}{\sf\small$\blacktriangleright$\textit{#2}$\blacktriangleleft$}}}
	
 }{
 	\newcommand{\nbc}[3]{}
 	
 }

\usepackage[most]{tcolorbox}
\ifthenelse{\boolean{showedits}}
{
  \newtcolorbox{inserted}{%
       title=Inserted text:,
       colframe=blue,colback=blue!5!white,
       breakable,
       leftrule=0mm, 
       bottomrule=0mm,
       rightrule=0mm,
       toprule=0mm,
       arc=0mm, outer arc=0mm,
       oversize
  }
  \newtcolorbox{deleted}{%
       title=Deleted text:,
       colframe=red,colback=red!5!white,
       breakable,
       leftrule=0mm, 
       bottomrule=0mm,
       rightrule=0mm,
       toprule=0mm,
       arc=0mm, outer arc=0mm,
       oversize
  }
  \newtcolorbox{refactored}{%
       title=Rewritten text:,
       colframe=blue,colback=red!5!white,
       breakable,
       leftrule=0mm, 
       bottomrule=0mm,
       rightrule=0mm,
       toprule=0mm,
       arc=0mm, outer arc=0mm,
       oversize
  }
}{

}
\newboolean{isblinded}
\setboolean{isblinded}{true}
\ifthenelse{\boolean{isblinded}}
{\newcommand\blind[1]{BLINDED\xspace}}
{\newcommand\blind[1]{#1\xspace}}

\newcommand{\commented}[1]{}

\newcommand{\etal}{\emph{et al.}\xspace}

\definecolor{source}{gray}{0.9}

\usepackage{draftwatermark}
\usepackage{transparent}
\SetWatermarkText{\transparent{0.5}\fontsize{50}{54}\selectfont \parbox[c][3cm][c]{\paperwidth}{\centering Preprint\\CSEE\&T 2024}}
\SetWatermarkScale{1}

\begin{document}

\title{
Gameful Introduction to Cryptography for Dyslexic Students
}

\author{
\IEEEauthorblockN{Argianto Rahartomo}
\IEEEauthorblockA{
Technische Universität Clausthal\\Germany\\
\href{https://orcid.org/0000-0002-9592-0023}{ORCID: 0000-0002-9592-0023}
}
\and
\IEEEauthorblockN{Harpreet Kaur}
\IEEEauthorblockA{
Technische Universität Clausthal\\Germany\\
\href{https://orcid.org/0009-0004-5276-1618}{ORCID: 0009-0004-5276-1618}
}
\and
\IEEEauthorblockN{Mohammad Ghafari}
\IEEEauthorblockA{
Technische Universität Clausthal\\Germany\\
\href{https://orcid.org/0000-0002-1986-9668}{ORCID: 0000-0002-1986-9668}
}
}

\maketitle

\begin{abstract}
Cryptography has a pivotal role in securing our digital world. 
Nonetheless, it is a challenging topic to learn.  
In this paper, we show that despite its complex nature, dyslexia—a learning disorder that influences reading and writing skills—does not hinder one's ability to comprehend cryptography. 
In particular, we conducted a gameful workshop with 14 high-school dyslexic students and taught them fundamental encryption methods.
The students engaged well, learned the techniques, and enjoyed the training. 
We conclude that with
a proper approach, dyslexia cannot hinder learning a 
complex subject such as cryptography.

\end{abstract}

\IEEEpeerreviewmaketitle

\section{Introduction}
\label{sec:introduction}
Dyslexia, identified by the World Federation of Neurology in 1968, is a specific learning disorder of neurological origin, characterized by challenges in accurate and fluent word recognition, spelling, and decoding abilities~\cite{miles_living_1999}.
Dyslexic individuals often encounter challenges in reading and writing that are disproportionate to their intellectual capabilities, leading to misconceptions about their academic potential~\cite{wuyanqi2022}. 
Indeed, the typical reliance on textual and numerical teaching methods can be a hurdle for individuals with dyslexia~\cite{linda2023}.
Unfortunately,
dyslexics are blamed for their poor academic performance, which can lead to psychological issues such as poor self-image or low self-esteem~\cite{Zdravkova_2022}. 
Consequently, 
previous work concluded dyslexic students underperform in Mathematics~\cite{xinwei2013}, and dyslexic students are underrepresented in Science, Technology, Engineering, and Mathematics (STEM) fields~\cite{dunn2012}. 
Nonetheless,
recent studies have shown that online and computer-based solutions are effective in offering good learning experiences to dyslexic people~\cite{kautish_existing_2021,dimitriadou_online_2023}.

In the pursuit of providing an inclusive education, in this paper, we show that dyslexic students can learn a complex topic when provided with an appropriate approach.
In particular,
we offered a gameful workshop to introduce cryptography techniques and provided an online learning platform where participants could practice each technique.
There were 14 dyslexic participants from high schools, and the aim was to teach them encryption methods such as the Caesar, Vigenère, and Playfair ciphers, which establish the foundation for more advanced cryptographic principles.
We chose cryptography subject for two main  reasons:
(1) It is pivotal to secure the digital world and the curriculum of undergraduate programs often includes this subject.
(2) It is a challenging topic due to its complex mathematical base and abstract principles~\cite{song2009}. 
We adopted a gamified approach to make the learning experience more enjoyable for students, and we used a multisensory approach (a combination of visual, auditory, and tactile stimuli) to improve our communications.

In summary, 
we confirm that dyslexia is not a barrier to success in technical disciplines.
We show that dyslexic students can learn a complex subject such as cryptography which is challenging to learn for everyone.
We hope that this study 
empower dyslexic students with the confidence required to pursue their interests, regardless of the topic's complexity.

The workshop materials, CryptoLexia's code, demos, and experiment data are publicly available.\footnote{https://bit.ly/cseet24dyslexia}

The remainder of this paper is structured as follows.
Section~\ref{sec:background} introduces the cryptography topics that we covered in our workshop.
Section~\ref{sec:method} describes our gamful teaching method.
Section~\ref{sec:result} presents the results, and Section~\ref{sec:threat} discusses the potential threats to validity of our study.
Section~\ref{sec:literaturereview} makes an overview of related work.
Section~\ref{sec:conclusion} concludes this paper.

\section{Background}
\label{sec:background}

Cryptography has a pivotal role in providing secure digital solutions.
It involves techniques to encode data (i.e., \textit{encryption}) so that only authorized parties can decode it (i.e., \textit{decryption}), therefore maintaining data confidentiality.
Despite its importance, developers often struggle with the correct adoption of cryptography APIs~\cite{Hazhirpasand18, Hazhirpasand2021, Hazhirpasand2021b}, leading to widespread misuses~\cite{Hazhirpasand2020}.
This highlights the need to motivate students to learn cryptography and prepare them for this topic.

Three foundational encryption techniques are Caesar, Vigenère, and Playfair ciphers~\cite{van1999fundamentals}.
They introduce basic principles and techniques that underlie more complex and modern cryptographic methods.

\subsection{Caesar Cipher}
The Caesar cipher is one of the simplest encryption techniques. Named after the historical figure Julius Caesar, who lived in the 1st century BCE, it was used to communicate secretly with his correspondents.
Technically, this method involves shifting the alphabet (\textit{plaintext}) by a predetermined number of positions (\textit{key}) agreed upon by both parties.
For example, consider the plaintext ``\texttt{all good things}'' with a key of 7.
%
Each letter in the plaintext will be shifted seven positions (for example, ``a'' $\rightarrow$ ``h''), and the final ciphertext will be ``\texttt{hss nvvk aopunz}''.


%


%
The Caesar cipher provides limited security due to the fixed number of shifts and the use of a 26-letter alphabet.

\subsection{Vigenère Cipher}

The Vigenère cipher, named after Blaise de Vigenère in the 16th century also involves shifting the alphabet, but unlike the Caesar cipher, the shift is not static and depends on a secret key.
%
%
For example, consider the plaintext ``\texttt{all good things}'' and ``\texttt{secure}'' as the secret key.
Firstly, we should repeat the secret key until it matches the number of characters in the plaintext. Hence, we obtain a new secret key i.e., ``\texttt{securesecures}''.
Next, we create tuples of characters that are at the same index in plaintext and secret key.
For instance, (a, s) at the first index, (l, e) at the second index, etc.
Then, we use the Vigenère Table to find the intersection of the two characters in each tuple.\footnote{\url{https://www.boxentriq.com/code-breaking/vigenere-cipher\#facts}}
Particularly, we search for a character in the x-axis if it belongs to the plaintext, and we search for a character in the y-axis if it belongs to the key.
Hence, for (a, s) the intersection is ``s'', for (l, e) it is ``p'', etc.
The intersections for every two corresponding characters result in the ciphertext which would be ``\texttt{spn afsw xjcekk}''.

Compared to the Caesar cipher, the Vigenère cipher produces a ciphertext with a more varied and less predictable distribution of letters.
This reduces the likelihood of guessing the original text based on letter frequency, making it more complex than the Caesar cipher.
However, the Vigenère cipher is vulnerable to frequency analysis attacks if the secret key is too short or repetitive~\cite{asmitha2024}.

\subsection{Playfair Cipher}

The Playfair cipher, developed by Charles Wheatstone in the 19th century, is named after Lord Playfair, who promoted the use of this cipher.
This cipher involves encrypting pairs of letters (digraphs) using a 5x5 matrix constructed from the secret key.
For instance, consider ``\texttt{all good things}'' as the plaintext and ``\texttt{secure}'' as the secret key.

\begin{table}
\caption{Playfair cipher matrix for key ``secure''}
\centering
\begin{tabular}{|c|c|c|c|c|}
\hline
s & e & c & u & r \\ \hline
a & b & d & f & g \\ \hline
h & i & k & l & m \\ \hline
n & o & p & q & t \\ \hline
v & w & x & y & z \\ \hline
\end{tabular}
\label{tab:playfair}
\end{table}

In the first step, we must remove any duplicate letters from the secret key. 
For ``\texttt{secure}'' this results in ``\texttt{secur}''.
We fill in the 5x5 cipher matrix with this key and then complete the remaining spaces with the other letters of the alphabet in order.\footnote{There are different matrix sizes, but the traditional Playfair cipher uses a 5x5 matrix. In this matrix, the letter ``j'', one of the least common letters in English, is treated the same as ``i'', which results in ``j'' exclusion.}
Table~\ref{tab:playfair} shows the result.

In the second step, we construct digraphs.
We begin by dividing the plaintext into pairs of letters, moving left to right. It becomes ``\texttt{al lg oo dt hi ng s}''.
We then apply two rules.
(1) If a pair has the same letter twice, such as ``\texttt{oo}'', we insert an ``\texttt{x}'' between them, changing ``\texttt{oo}'' to ``\texttt{ox}''.
(2) If the plaintext ends with an odd number of characters, we add an ``\texttt{x}'' to the last character, ensuring the plaintext length becomes even, such as changing ``\texttt{s}'' to ``\texttt{sx}''.
Hence, we obtain the pairs listed below, which comprise the digraphs ``\texttt{al lg ox od th in gs}''.

\begin{itemize}
            \item ``\texttt{al}'' $\rightarrow$ ``\texttt{al}''
            \item ``\texttt{lg}'' $\rightarrow$ ``\texttt{lg}''
            \item ``\texttt{oo}'' $\rightarrow$ ``\texttt{ox}''
            \item ``\texttt{dt}'' $\rightarrow$ ``\texttt{od}''
            \item ``\texttt{hi}'' $\rightarrow$ ``\texttt{th}''
            \item ``\texttt{ng}'' $\rightarrow$ ``\texttt{in}''
            \item ``\texttt{s}'' $\rightarrow$ ``\texttt{gs}''
\end{itemize}

Finally, we use the cipher matrix to encrypt each digraph. 
In particular, there are three rules.

    \begin{itemize}
        \item If both letters are in the same column, select the letter directly below each one (wrapping to the top if necessary). For instance, ``\texttt{sh}'' $\rightarrow$ ``\texttt{an}''.
        
        \item If both letters are in the same row, select the letter immediately to the right of each one (wrapping to the leftmost position if necessary). For instance, ``\texttt{hk}'' $\rightarrow$ ``\texttt{il}''.
        
        \item If neither of these conditions is met, form a rectangle with the two letters and take the letters at the horizontal opposite corners of the rectangle. For instance, ``\texttt{ed}'' $\rightarrow$ ``\texttt{cb}''.
    \end{itemize}

%

Applying these rules to our digraphs, the ciphertext would be ``\texttt{fhm fpwpb nmhoar}''.

Although Playfair cipher is more complex than the other two ciphers, it is susceptible to digraph frequency attacks~\cite{maha2020}.
%


\section{Teaching Method}
\label{sec:method}

We held an online workshop via the Zoom platform, with 14 voluntary participants identified as dyslexic. 
They were from seven high schools in India (7 to 12 grades) and had a gender distribution of 9 males and 5 females.
The decision to hold an online workshop was based on practicality and accessibility, allowing for broader participation without geographical limitations. 
The participants' selection was based on our established connections with these schools. 
To ensure participants' privacy and comfort, we did not record the workshop sessions, did not collect any sensitive/personal data, and ensured that the learning activities occurred in a safe and controlled environment. 
We reviewed our process with representatives (self-volunteered teachers) from each school and ensured that all activities and interactions were designed to avoid the use or collection of participants' data.

We started the workshop with a general introduction to cryptography and the encryption and decryption procedures. 
We then provided detailed explanations of each encryption technique.
We told stories to spark student interest and aid the flow of the workshop.
We utilized Microsoft PowerPoint to present the content, included supportive visual aids, used mind maps to enhance memory retention for cryptography concepts, and adjusted the learning pace based on feedback from the participants.

%
%

We covered Caesar, Vigenère, and Playfair ciphers in this workshop. 
We explained each technique in detail and played a game with participants to practice every technique together.
We proceeded to the next cipher only when there was no question.

We concluded the workshop by presenting ``CryptoLexia'', an open-source web-based learning platform to practice each technique and acquire hands-on experiences.
%
We developed CryptoLexia using the VueJS framework,\footnote{
https://vuejs.org} ``Open Dyslexic'' typography,\footnote{
https://opendyslexic.org} and ``Speecify'' text-to-speech tool.\footnote{https://speechify.com}
In addition, we relied on Kahoot! to create engaging and interactive quizzes.\footnote{https://kahoot.com} 

We aimed to provide a gameful experience for students and incorporated the following gamification elements into our teaching approach.

\subsection{Levels}

The levels are designed to represent different encryption techniques, progressively increasing in complexity. The game starts with Level 1, which covers the Caesar cipher. Level 2 explores the Vigenère cipher, and Level 3, the most challenging level, focuses on the Playfair cipher.

\subsection{Storytelling}

We told a story to explain the necessity of ciphers and engage students in secret protection. 

Once upon a time, there was a war in an occupied country. They knew that a new attack would happen, but the enemy had communicated the exact attack plan using ciphers so that no one could understand their secrets.
One of the soldiers happened to find a suspicious letter written in Caesar Cipher code. The soldier showed the letter to the cryptography team, who tried to decrypt the message.
They succeeded and discovered the name of an old palace in the north.
Troops went to the palace, searched for more evidence, and found new letters encrypted with Playfair cipher. 
They decoded the letters and discovered the name of a spy in the country.
They arrested him and found an image of the country in his pocket with some text encoded in Vigenère cipher.
The team decrypted the text and found the exact timing of the attack.


%


\subsection{Interaction}

The game includes a text-to-speech feature that helps players read and engage with the content. In the next version, we plan to add an alternative input method, such as speech. Furthermore, the game provides hints at each level to assist students in recalling relevant information.

\subsection{Feedback}

Participants could attempt the game repeatedly and without restriction for one week.
We did not enable recording the ``number of attempts'' nor ``time to completion'' to ensure that the learning experience is stress-free.

Players receive a score for each correct answer. In general, as they progress in the game, the questions become more challenging, and so the score for each question increases. 
The game features a scoreboard where players are ranked among their peers, enabling students to track their progress and compare their points. This competitive aspect aims to motivate students to improve their performance.

\section{The Result}
\label{sec:result}


In general, all participants completed the challenges that were offered in the CryptoLexia platform. 
We surveyed the participants and collected their feedback one week after the workshop.
They responded to ten questions listed in Table~\ref{tab:post-testinterviews}.
%
%

\begin{table}[ht]
    \centering
    \caption{The questionnaire}
    \label{tab:post-testinterviews}
    \begin{tabular}{lp{7cm}} \toprule
        \textbf{\#} & \textbf{Question} \\ \midrule
        \multirow{2}{*}{Q1}&Did you enjoy this workshop?\\
        &Options: Yes, No, or Not Sure\\[2pt]
        \multirow{2}{*}{Q2}&How would you rate your learning experience?\\
        &Likert scale 1: Worst to 5: Best \\[2pt]
        \multirow{2}{*}{Q3}&Was the subject clearly communicated?\\
        &Options: Very much, Somewhat, Undecided, Not Really\\[2pt]
        \multirow{2}{*}{Q4}&Do you prefer online modules or in-person activities?\\
        &Options: Online, Offline\\[2pt]
        \multirow{2}{*}{Q5}&Did you feel supported throughout this workshop?\\
        &Options: Very much, Somewhat, Undecided, Not Really\\[2pt]
        \multirow{2}{*}{Q6}&The learning materials (content, visual elements, mind maps, etc.) were effective.\\
        &Likert scale 1: Strongly disagree to 5: Strongly agree \\[2pt]
        \multirow{2}{*}{Q7}&The content met my learning expectations.\\ 
        &Likert scale 1: Strongly disagree to 5: Strongly agree \\[2pt]
        \multirow{2}{*}{Q8}&I recommend this workshop.\\ 
        &Likert scale 1: Strongly disagree to 5: Strongly agree \\ [2pt]
        \multirow{2}{*}{Q9}&The workshop was relevant.\\ 
        &Likert scale 1: Strongly disagree to 5: Strongly agree \\ [2pt]
        \multirow{2}{*}{Q10}&The subject was challenging to understand.\\ 
        &Likert scale 1: Strongly disagree to 5: Strongly agree \\\bottomrule 
    \end{tabular} 
\end{table}

%

The survey results were positive as well.
Notably, the response to the game's enjoyment was unanimously positive, with all participants indicating their pleasure. They rated the game level's difficulty consistently high, predominantly scoring it as a 4 or 5 out of 5. 
A preference for online modules over in-person activities emerged, suggesting that they appreciate an effective digital learning platform.
Participants felt well-supported throughout the training, with a strong sense of agreement on the support provided. Learning aids such as workshops and mind maps were highly valued. The majority would recommend the game to a friend, reflecting its perceived value and appeal. 
However, opinions on the course relevance, with four neutral and two disagreeing responses, may be due to a lack of cryptography topics in the curriculum of high-school studies, and their limited knowledge about its applications in the real world.

\section{Threats to Validity}
\label{sec:threat}



Our study's integrity may be influenced by selection and self-selection biases, where the participants who volunteered might not represent the broader dyslexic population. We endeavored to mitigate this by reaching out to diverse educational settings.
Nonetheless, all participants were from the same state in India, which may not represent dyslexic students with other demographics.

Responses provided by participants could be subject to social desirability bias, potentially shaping their answers to align with perceived expectations. To lessen this effect, we assured anonymity and stressed the importance of honest feedback in our instructions, encouraging candidness without fear of judgment or repercussion.

The reliability of our measurement instruments is important, as inconsistent tools could yield unreliable 
data. 
We exercised the session and sought peer feedback on our tools to enhance their consistency and dependability.

Finally, 
it is necessary to obtain a more comprehensive understanding of how our approach specifically benefits dyslexic students, as well as identify any advantages they may have over traditional or other alternative teaching methods.
Therefore, future studies should include control groups such as a group of non-dyslexic students or a group receiving an alternative intervention.

\section{Related Work}
\label{sec:literaturereview}

Kilhoffer \etal~\cite{kilhoffer2023} investigated the teaching methods for topics such as cybersecurity and AI ethics among high school students in the United States (K-12). The research sample included 16 high school teachers and 11 students. They aimed to find out whether the instructional approaches employed by teachers effectively communicate these complex subjects to students. The findings showed that interactive discussions and gamification were particularly effective teaching strategies.
Interestingly, 
the non-STEM (i.e., social science and literature) teachers more frequently covered topics related to cybersecurity and AI ethics than STEM teachers.
This trend was supported by the belief among social science teachers that technological ethics are crucial for democracy and by literature teachers who emphasized the importance of critical analysis of information presented to us. 


Lang \etal~\cite{lang2023} explored BrailleBuddy, a digital user interface meant to assist visually impaired children in learning Braille, which differed from non-digital learning approaches such as tactile books or Braille-labelled objects. This device combines microcontroller and platelets. The study aimed to examine whether visually impaired children could independently learn Braille using BrailleBuddy and evaluated their experience. The device was developed iteratively with feedback from four experts and tested with 11 blind participants, showing high effectiveness in assisting Braille learning. The study highlighted the potential of digital tools in promoting inclusive education. However, it noted some limitations for Attention Deficit Hyperactivity Disorder (ADHD) people, who may struggle to focus during BrailleBuddy learning sessions. Furthermore, voice modules from the microcontroller occasionally caused participant misunderstandings, prompting consideration of more effective audio aids like Speechify to assist dyslexic students in comprehending educational information.

Niklaus \etal~\cite{niklaus2023} studied how a digital reading software program, specifically a digital reading ruler, can help people with dyslexia. The study involved 177 participants, 91 of whom were dyslexic, and assessed the tool's usefulness in boosting reading speed. The findings showed that using a digital reading ruler helps dyslexics read faster. 

Alsobhi \etal~\cite{study001} introduced DAEL (Dyslexia Adaptive E-Learning) Framework. It integrates adaptive learning techniques with a keen understanding of dyslexic learners' needs, offering a personalized and effective learning experience. DAEL Framework considers four perspectives of Dyslexic learners: presentation, hypermediality, accessibility and acceptability, and user experience. The framework has been recognized for its potential in not only facilitating a more inclusive learning environment but also in contributing to our understanding of how technology can be leveraged to address specific educational challenges faced by students with dyslexia.

Minoofam \etal~\cite{study002} introduced RALF, an Adaptive Reinforcement Learning Framework, specifically designed to enhance the educational experience of dyslexic students. RALF utilizes the principles of reinforcement learning to create a dynamic and responsive learning environment, one that adapts to the unique learning pace and style of each dyslexic learner. The framework's underlying strength lies in its ability to continuously adjust educational content. 

Saeed \etal~\cite{Saeed2022} conducted a systematic literature review to identify the use of game elements and game-based intervention methods to enhance the learning capabilities of dyslexic patients. 
They collected 42 relevant papers from 2011 to 2020, and they found that word exercise-based games are the most common game type.

In summary, game-based interventions are proven to be effective therapeutic aids for learning difficulties. 
We also adopted a gameful approach in our workshop to facilitate teaching and enhance students' understanding.

\section{Conclusion}
\label{sec:conclusion}
This study demonstrates that with a tailored teaching approach, we can effectively engage dyslexic individuals in a complex topic such as cryptography. 
The positive outcomes observed suggest that dyslexia does not hinder the ability to learn difficult technical topics. 
Dyslexics should not underestimate their ability to learn; instead, they should seek the right pedagogy that aligns with their needs.

\section*{Acknowledgment}

We are thankful for the support of the high school delegates who facilitated the conduct of this study.

\bibliographystyle{IEEEtran}
\bibliography{main}

\end{document}